\documentclass[10pt,article]{IEEEtran}
\usepackage[linesnumbered,ruled,vlined]{algorithm2e}
\usepackage{srcltx,pdfsync}
\usepackage{amsmath}
\usepackage{amsfonts}
\usepackage{nopageno}
\usepackage{amssymb}
\usepackage{graphicx}
\usepackage{balance}
\usepackage{epstopdf}
\usepackage{cite}
\usepackage{url,ifthen}
\usepackage{caption}
\usepackage{subcaption}
\usepackage{mathrsfs}
\usepackage{amsthm}
\usepackage[dvipsnames,usenames]{color}
\usepackage[colorlinks, linkcolor=Blue, filecolor =Red,
citecolor=RoyalPurple]{hyperref}
\usepackage[lmargin=0.76in,rmargin=0.76in,tmargin=0.73in,bmargin=0.79in]{geometry}

\newtheorem{prop}{Proposition}

\newtheorem{assumption}{Assumption}
\newtheorem*{proof-non}{Proof}
\newtheorem{remark}{Remark}

\def\Expect{{\sf E}}
\def\Prob{{\sf P}}
\def\vecOfG{{\sf G}}
\def\vecOfQoS{{\sf b}}
\def\setOfPSD{\ensuremath{\mathcal{S}}}

\def\discFreq{{\sf \ensuremath{\Omega}}}
\def\eqdef{\ensuremath{:=}}

\def\numLoads{\ensuremath{n}}

\def\netDem{\text{\scriptsize ND}}
\def\loadDem{\text{\scriptsize BA}}
\def\PSD{\text{SD}}
\def\COP{0}
\def\INT{\text{\scriptsize int}}

\def\textLow{\text{\scriptsize Low}}
\def\textHigh{\text{\scriptsize High}}

\def\G{\ensuremath{\mathcal{G}}}
\def\R{\ensuremath{\mathbb{R}}}

\newlength{\noteWidth}
\long\def\notes#1{\ifinner
           {\footnotesize #1}
           \else
           \marginpar{\parbox[t]{\noteWidth}{\raggedright\footnotesize #1}}
       \fi\typeout{#1}}
\def\pb#1{\notes{pb: {\color{red}{#1}}      }}

\usepackage{xparse}

\ExplSyntaxOn
\cs_new:Nn \pb_prop_gset_bykeys:Nn
{
	\prop_clear:N \l__pb_temp_prop
	\keys_set:nn { pb/propbykey } { #2 }
	\prop_gset_eq:NN #1 \l__pb_temp_prop
}
\cs_generate_variant:Nn \pb_prop_gset_bykeys:Nn { c }

\keys_define:nn { pb/propbykey }
{
	unknown .code:n = \prop_put:Nxn \l__pb_temp_prop { \l_keys_key_tl } { #1 }
}

\cs_generate_variant:Nn \prop_put:Nnn { Nx }

\NewDocumentCommand{\setupcollaborator}{mm}
{
	\prop_new:c { g_collaborator_#1_prop }
	\pb_prop_gset_bykeys:cn { g_collaborator_#1_prop } { #2 }
}

\NewDocumentCommand{\selectcollaborator}{m}
{
	\prop_map_inline:cn { g_collaborator_#1_prop }
	{
		\tl_set:cn { ##1 } { ##2 }
	}
}
\ExplSyntaxOff

\setupcollaborator{GZlinux}
{
	NAME=GZ on his lab computer,
	DiCEbibPATH=../../../../../DiCElab_bib,
	JBbibPATH=../../../../../bibs,
}
\setupcollaborator{Austin}
{
	NAME=AC on his lab computer,
	DiCEbibPATH=../../bibFiles/dicelab_bib,
	JBbibPATH=../../bibFiles/bibs,
}
\setupcollaborator{PBmacbookair}
{
	NAME=Prabir Macbookair,
	DiCEbibPATH=/Users/pbarooah/Dropbox/Dropbox/DiCElab_bib,
	JBbibPATH=/Users/pbarooah/Dropbox/Dropbox/bibs,
}

\selectcollaborator{Austin}

\begin{document}
\bstctlcite{IEEEexample:BSTcontrol} 
\title{\vspace{6.4mm}A model-free method for learning flexibility capacity of loads providing grid support}
\author{
	\IEEEauthorblockN{Austin R. Coffman\IEEEauthorrefmark{1}$^,$\IEEEauthorrefmark{2} and Prabir Barooah\IEEEauthorrefmark{1}}  \vspace{-0.7cm}
	
	\thanks{\IEEEauthorrefmark{1} University of Florida}
	\thanks{\IEEEauthorrefmark{2} corresponding author, email: bubbaroney@ufl.edu.}
	\thanks{ARC and PB are with the Dept. of Mechanical and Aerospace Engineering, University of Florida, Gainesville, FL 32601, USA. The research reported here has been partially supported by the NSF through awards \#1646229 and \#1934322.}
}

\maketitle
\thispagestyle{empty}
\begin{abstract}
Flexible loads are a resource for the Balancing Authority (BA) of the future to aid in the balance of power supply and demand. In order to be used as a resource, the BA must know the capacity of the flexible loads to vary their power demand over a baseline without violating consumers' quality of service (QoS). Existing work on capacity characterization is model-based: They need models relating power consumption to variables that dictate QoS, such as temperature in case of an air conditioning system. However, in many cases the model parameters are not known or difficult to obtain. In this work, we pose a data driven capacity characterization method that does not require model information, it only needs access to a simulator. The capacity is characterized as the set of feasible spectral densities (\PSD s) of the demand deviation. The proposed method is an extension of our recent work on SD-based capacity characterization that was limited to linear time invariant (LTI) dynamics of loads. The method proposed here is applicable to nonlinear dynamics.  Numerical evaluation of the method is provided, including a comparison with the model-based solution for the LTI case. 

\end{abstract}
\section{Introduction}

The future of the power grid is green: an increased reliance on renewable generation. This can pose a challenge for Balancing Authorities (BAs) since renewable generation is volatile. To ensure balance of power in the presence of this volatility, BAs need additional sources of energy storage. Apart from batteries, a new resource has been the subject of much investigation for its ability to provide battery-like service: flexible loads.
 
Most loads have some flexibility in power demand: they can deviate their power demand from a baseline value without violating their quality of service (QoS). The BA would request this demand deviation, termed the \emph{reference signal}, so to help balance the grid. The baseline consumption is then power consumption in absence of any requests from the BA. Examples of flexible loads include pumps for pool cleaning~\cite{ChenDistributedIMA:2017} and agricultural purposes~\cite{KiedanskiExploitingSGC:2019}, TCLs~\cite{KhurramIdentificationEPSR:2020}, and HVAC equipment~\cite{CaiLaboratoryENB:2019}.  

If the grid operator expects the flexible loads to track the reference signal accurately, then the reference must not cause the loads to violate their QoS. From the viewpoint of the grid operator, flexible loads not tracking a reference makes them appear unreliable. From the viewpoint of the load, reference signals that continually require QoS violation provide incentive for loads to stop providing grid support. In either case, avoidance of the above scenarios is paramount to the long term success of grid support from flexible loads. That is, reference signals must be designed to respect the \emph{capacity} of the collection of flexible loads.

Informally, the capacity of of a flexible load represents limitations in its ability to track a demand deviation reference signal due to QoS requirements at the individual loads. Consequently, a key step in determining the capacity is relating the QoS requirements to requirements for the reference signal for demand deviation. This is a challenging task; various approaches have been proposed in recent years~\cite{hao_aggregate:2015,haowuliayan:2017,coffman2019aggregate:arxivACC,buildFlex:yin:2016,HaoKalsi:CDC:15,HughesPoolla:HICCS:15,ChakrabortyDataBookChap:2020,AmasyaliMachineISGT:2020}. A popular approach is to develop ensemble level necessary conditions~\cite{hao_aggregate:2015,coffman2019aggregate:arxivACC}. Reference signals that satisfy these conditions ensure the ability of all loads in the collection to satisfy QoS while tracking the reference. Other approaches include geometry based characterizations~\cite{KunduKalsi:PSCC:18}, characterizations through distributed optimization~\cite{Lin_ACC2018}, and characterizations that approximate the Minkowski sum of individual load's ``resource polytopes''~\cite{MullerAggregationCDC:2015,BarotConciseIJEPES:2017,NazirInnerCDC:2018}.

\begin{figure} [t]
	\centering
	\includegraphics[width=1\columnwidth]{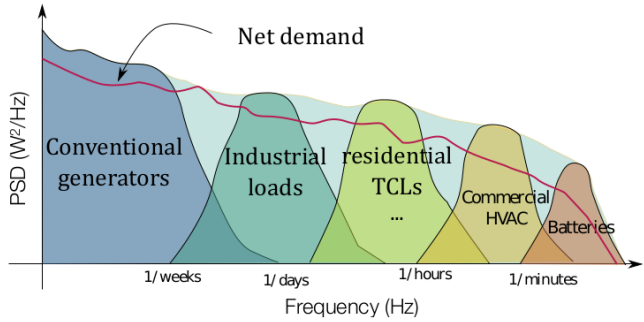}
	\caption{An example spectral allocation of resources to meet the grids needs.}
	\label{fig:Resource_Power}
\end{figure}

In addition to the previous references, there is an emerging methodology of characterizing the capacity as constraints on the statistics of the reference signal~\cite{barbusmey:2015,WangFrequencyISGT:2020,CoffmanSpectral_ACC:2020}, rather than the reference signal itself. Most commonly, as constraints on the \emph{spectral density} (SD) of the reference signal~\cite{WangFrequencyISGT:2020,CoffmanSpectral_ACC:2020}. Elaborating, these methods aim to precisely quantify the  regions shown in Figure~\ref{fig:Resource_Power} based on the QoS of the loads considered. One particular advantage of these characterizations is that they are suitable for long term resource allocation. That is, such a characterization can answer questions such as: how many flexible loads will a BA require if it invests in 10\% more solar? Contrarily, if the capacity of the flexible loads is characterized in the time domain, the BA would then have to predict many months in advance solar and flexible load trajectories to see if the flexible loads can deliver. 

Irrespective of statistical or time domain characterizations, many of the listed works have one thing in common: they are model based. Meaning, they need a model that relates demand deviation of the flexible load(s) to QoS. The computed capacity of the load(s) depend on the model/model parameters. In addition, most of the existing methods require the models to be linear time invariant (LTI). An LTI model may be inappropriate for certain loads. Even if a linear model is appropriate, the parameters of the flexible load model are typically unknown or require estimation from experimental data or high-fidelity simulations. In the spirit of model free control, one might wonder then, is it possible to directly estimate a characterization of flexible load capacity from data?

In this work, we develop a data-driven (model free) framework to determine flexible load capacity directly from data. That is, instead of relying on model knowledge it rather relies on access to a simulator to determine the capacity. This framework builds off of our past work~\cite{CoffmanSpectral_ACC:2020}, where we characterize the capacity of flexible load(s) as constraints on the \PSD\ of their power deviation. In the past work, to obtain a \PSD\ we set up an optimization problem: the BA projects its needs (roughly, the \PSD\ of net demand, e.g., shown in Figure~\ref{fig:Resource_Power}) onto the constraint set of feasible of \PSD's. In~\cite{CoffmanSpectral_ACC:2020}, the models relating demand deviation to QoS variables were assumed LTI, and it was shown how to solve the optimization problem for that case.

We depart from our previous work here and venture into new territory by showing how to solve the projection problem from our past work without having access to model information, as long as we have the ability to generate data from a simulator. The key insight that allows for this model free construction is the choice of approximation architecture when discretizing the infinite dimensional optimization problem. This data driven framework also has the ability to accommodate non-linear flexible load models, whereas our past work~\cite{CoffmanSpectral_ACC:2020} was for linear models only.  


We validate our data driven framework in simulation examples. First, we compare the proposed data driven framework to the model based framework of our past work~\cite{CoffmanSpectral_ACC:2020}. We then use our framework to estimate the capacity of a flexible load with a non-linear model. 

The paper proceeds as follows. In Section~\ref{sec:priorWork} we introduce the method from our prior work. In Section~\ref{sec:sepCase} we introduce our data-driven method. Numerical experiments are conducted in Section~\ref{sec:numExp} and we conclude in Section~\ref{sec:conc}

\section{Spectral Characterization of QoS Constraints} \label{sec:priorWork}
The symbol $t$ is used to denote the continuous time while $k$ is used to denote a discrete time index. The sampling interval is $\Delta t$.

\subsection{Deterministic QoS constraints} \label{sec:indLoadModel}
Denote by $P[k]$ the power consumption of a flexible load at time index $k$, and let $P^b[k]$ its  baseline demand. The demand deviation is $\tilde{P}[k] \eqdef P[k] - P^b[k]$. The load provides grid support service by controlling the deviation $\tilde{P}[k]$ to track a desired deviation signal, called a reference, while maintaining its own QoS. The first QoS constraint is simply an actuator constraint:
\begin{align} \label{eq:QoSOne}
&\text{QoS-1:} \quad \left|\tilde{P}[k]\right| \leq c_1, \quad \forall \ k,
\end{align}
where the constant $c_1$, the maximum possible deviation of power consumption, depends on the rated power and the baseline demand. Second, define the demand increment $\tilde{P}_\delta[k] \eqdef \tilde{P}[k]-\tilde{P}[k-\delta]$, where $\delta>0$ is a predetermined (small) integer time interval. The second constraint is a ramping rate constraint:
\begin{align}
\label{eq:QoS-increment}
\text{QoS-2:} \quad \left|\tilde{P_\delta}[k]\right| \leq c_2, \quad &\forall \ k.
\end{align}
Third, define the additional energy use during any integer time interval of length $T$:
\begin{align} \label{eq:engDevDyn}
\tilde{E}[k] = \sum_{\sigma = k-T+1}^{k}\tilde{P}[\sigma].
\end{align}
The third QoS constraint is that
\begin{align}\label{eq:QoS_energy}
\text{QoS-3:} \quad \left|\tilde{E}[k]\right| \leq c_3, \quad &\forall \ k.
\end{align}
The parameter $T$ in~\eqref{eq:engDevDyn} can represent the length of a billing period. Ensuring~\eqref{eq:QoS_energy} ensures that the energy consumed during a billing period close to the nominal energy consumed, although it is stronger than what is necessary. 

To define the fourth and last QoS constraint, we associate with the VES system a \emph{storage variable} $\tilde{x}[k]$ that is related to the demand deviation, and impose the constraint
\begin{align}\label{eq:QoS_Thermenergy}
\text{QoS-4:} \quad \left|\tilde{x}[k]\right| \leq c_4, \quad &\forall \ k.
\end{align}

\subsubsection{Understanding QoS-4} To understand the storage variable, imagine a flexible HVAC system providing VES. We first present a model of the HVAC systems internal temperature $T_z$ in continuous time, as it is more naturally presented in this setting. This model is:
\begin{align} \label{eq:thermDyn}
C\frac{dT_z(t)}{dt} = \frac{1}{R}\left(T_a(t) - T_z(t)\right) + \dot{q}_{\INT}(t) + Q(t), 
\end{align}
where $C$ and $R$ are thermal capacitance and resistance, $T_a(t)$ is the ambient temperature, and $\dot{q}_{\INT}(t)$ is an exogenous disturbance. The quantity $Q(t)$ is the rate of heat delivered to the building by the HVAC system (negative if cooling).

We consider two models, one linear and the other nonlinear for relating the electrical power deviation to the indoor temperature. In both cases a temperature deviation will play the role of the storage variable $\tilde{x}[k]$.

\paragraph{Linear model}
Suppose $Q(t) = -\eta_{0}P(t)$ where $\eta_{0}$ is the coefficient of performance (COP) under design conditions. In general, the baseline power consumption for a HVAC system is the value $P^b(t)$ that keeps the internal temperature of the load at a fixed value $\bar{T}$, which for~\eqref{eq:thermDyn} is
\begin{align} \label{eq:baseLinModel}
P^b(t) = -\frac{\left(T_a(t)-\bar{T}\right)}{\eta_{0} R} - \frac{\dot{q}_{\INT}(t)}{\eta_{0}}.
\end{align} 
Since we are concerned with the flexibility in the load, we linearize~\eqref{eq:thermDyn} about the thermal setpoint $\bar{\theta}$ and the baseline power $P^b(t)$ yielding,
\begin{align}  \label{eq:devModel}
\dot{\tilde{T}}_z(t) = -\gamma\tilde{T}_z(t) + \beta\tilde{P}(t), \quad \gamma = \frac{1}{RC}, \quad \beta = \frac{\eta_{0}}{C},
\end{align} 
where $\tilde{T}_z\triangleq T_z(t) - \bar{T}$ is the internal temperature deviation and $\tilde{P}$ is as defined at the beginning of this Section. The corresponding discrete-time dynamic model relating $\tilde{P}[k]$ to $\tilde{T}_z[k]$ is
\begin{align}  \label{eq:storage-LTI}
\tilde{T}_z[k+1] = a\tilde{T}_z[k] + b\tilde{P}[k],
\end{align}
(where $a =e^{-\gamma \Delta t}$ and $b = \beta\int_{0}^{\Delta t}e^{-\gamma \tau}d\tau$), which is also a first order linear time invariant (LTI) model.

\paragraph{Nonlinear model} A more realistic model is a COP that varies depending on the difference between indoor and outdoor temperature. When the HVAC system is providing cooling, the hotter the outside is compared to the indoors, the less efficient the HVAC system is in rejecting heat from the cooler indoor to the hotter outdoor~\cite{ASHRAE_handbook_fund:17}. Such a situation can be modeled as
\begin{align} \label{eq:nonLinCOPMod}
\eta(t) = \eta_{0} - \alpha_1 \big(T_a - T_z(t)\big) + \alpha_2.
\end{align}
The role of the constant $\alpha_2$ is to get $\eta(t) = \eta_{0}$ when $T_a$ and $T_z$ are both constant and equal to the values the HVAC system is designed for. The dynamic equation~\eqref{eq:thermDyn} with this model for the COP then becomes the following nonlinear ODE
\begin{align} \nonumber
C\dot{T}_z &= -\frac{1}{R}\big(T_a - T_z(t)\big) + \dot{q}_{\INT}(t)  \\ \label{eq:nonLinModel}
&+ \Big(\eta_0
- \alpha_1 \big(T_a - T_z(t)\big) + \alpha_2\Big)P(t).
\end{align}
The baseline for this model is the expression~\eqref{eq:baseLinModel}, except replacing $\eta_{\COP}$ with 	$\bar{\eta} = \eta_{\COP} - \alpha_1\Big(T_a - \bar{T}\Big) + \alpha_2.$ The corresponding ODE that relates the power deviation $\tilde{P}(t)$ to temperature deviation $\tilde{T}_z(t)$ is the following bilinear ODE:
\begin{align} \nonumber
 C\frac{d\tilde{T}_z(t)}{dt} &= -\frac{1}{R}\tilde{T}_z(t) + \Big(\eta_0 - \alpha_1 \big(T_a - \bar{T}_z(t)\big) + \alpha_2\Big)\tilde{P}(t) \\ \label{eq:ODE-nonlinear}
 &+\alpha \bar{P}\tilde{T}_z(t) + \alpha\tilde{T}_z(t)\tilde{P}(t).
 \end{align}
The derivation is tedious but straightforward, so it is omitted due to space limitations. The corresponding discrete time model - obtained with 1st order Euler backward discretization - is also a bilinear dynamic system.

\subsection{Mathematical Preliminaries}
In our prior work~\cite{CoffmanCharacterizingArxiv:2020}, we had developed a methodology that characterizes the capacity of a flexible load in the frequency domain. We briefly discuss this prior work here. Denote the power consumption of a flexible load as $\tilde{P}[k]$, where we model $\tilde{P}$ as a stochastic process. The mean and autocorrelation function for $\tilde{P}$ are,
\begin{align}
\mu_{\tilde{P}}[k] &\triangleq \Expect[\tilde{P}[k]], \quad \qquad \forall \ k, \\ 
R_{\tilde{P}}[s,k] &\triangleq \Expect[\tilde{P}[s]\tilde{P}[k]], \quad \forall \ s,\ k,
\end{align}
where $\Expect[\cdot]$ denotes mathematical expectation. In the past work, we made the following assumption about the stochastic process $\tilde{P}$.
\begin{assumption}\label{ass:WSSprocess}
	The stochastic process $\tilde{P}$ is wide sense stationary (WSS) with mean function $\mu_{\tilde{P}}[k] = 0$ for all $k$.
\end{assumption}
Under this assumption we have that the autocorrelation function will solely be a function of $\tau = s-k$. In this case, the autocorrelation function is an asymmetric Fourier transform pair with the Spectral Density:
\begin{align}\label{eq:WKT}
&R_{\tilde{P}}(\tau) = \frac{1}{2\pi}\int_{-\pi}^{\pi} S_{\tilde{P}}(\discFreq)  e^{j\discFreq\tau}d\discFreq, \ \text{and} \\ \label{eq:WKTTwo}
&S_{\tilde{P}}(\discFreq) = \sum_{\tau = -\infty}^{\infty} R_{\tilde{P}}[\tau]  e^{-j\discFreq\tau},
\end{align}
where $S_{\tilde{P}}(\discFreq)$ is the (power) Spectral Density (\PSD) of $\tilde{P}$, $\omega \in [-\pi,\pi]$ is the frequency variable, and $j$ is the imaginary unit. The above is based on the general definition of the \PSD\ of the signal ${\tilde{P}}$,
\begin{align} \label{eq:altDefPSD}
S_{\tilde{P}}(\discFreq) \triangleq \lim_{N \to \infty}\ \frac{1}{N}\Expect\bigg[\bigg|\sum_{k=1}^{N}\tilde{P}[k]e^{-j\discFreq k}\bigg|^2\bigg]
\end{align}
the equivalence of definitions~\eqref{eq:altDefPSD} and~\eqref{eq:WKTTwo} for a WSS process is the Wiener-Khinchin theorem. Since the mean function of $\tilde{P}$ is zero for all time we have
\begin{align} \label{eq:WKT_var}
\sigma^2_{\tilde{P}} = R_{\tilde{P}}(0) = \frac{1}{2\pi}\int_{-\pi}^{\pi} S_{\tilde{P}}(\discFreq) d\discFreq,	
\end{align}
that is, the variance $\sigma^2_{\tilde{P}}$ of $\tilde{P}$ is the integral of the (power) \PSD. To illustrate our method from prior work we will also make use of the Chebyshev inequality for a r.v. $X$:
\begin{align} \label{eq:chebIneq}
\Prob\big(\left|X - \mu_{\tiny X}\right| \geq C\big) \leq \frac{\sigma^2_X}{C^2}, \quad \forall \ C>0,
\end{align}
where $\Prob(\cdot)$ denotes probability. Another useful relation is the following:
\begin{prop}\label{prop:psdbackground}
If the input $x[k]$ to a linear time invariant system with frequency response $H(e^{j\discFreq})$ is WSS and has \PSD\ $\Phi_{x}$, then the output $y[k]$ is also WSS and its \PSD\ $\Phi_{y}$ is given by $\Phi_y(\discFreq) = \Phi_x(\discFreq)|H(e^{j\discFreq})|^2$.
\end{prop}

\subsection{Probabilistic QoS Constraints and \PSD-based Capacity Characterization} \label{sec:capsolvPSD}
Each QoS constraint potentially involves a distinct signal. In QoS-1, the signal is the power deviation $\tilde{P}[k]$ itself. In QoS-3, it is the storage variable $\tilde{x}[k]$. \emph{We denote by $Z_\ell[k]$ the signal relevant for the $\ell$-th QoS constraint.} In each QoS, the relevant signal is related to the power deviation $\tilde{P}[k]$, and we denote by $\G_\ell$ the (potentially dynamic) system that relates the input $\tilde{P}$ to the output $Z_\ell$.

Next we illustrate how to pose the QoS constraints as constraints on \PSD s. We start by considering the $\ell$-th QoS, which we re-formulate as
\begin{align}\label{eq:QoS3-probabilisic}
\Prob\left(\left| Z_{\ell}[k]\right| \geq c_\ell\right) \leq \varepsilon_\ell, \quad \forall k
\end{align}
where $\varepsilon_\ell \ll 1$ is the tolerance. From the Chebyshev inequality~\eqref{eq:chebIneq} and the equation~\eqref{eq:WKT_var} we have:
\begin{align} \label{eq:exampSuffCond}
\frac{1}{2\pi}\int_{-\pi}^{\pi} S_{Z_\ell}(\discFreq) d\discFreq \leq c_\ell^2\varepsilon_\ell \implies \Prob\left(\left|Z_{\ell}[k]\right| \geq c_\ell\right) \leq \varepsilon_\ell.
\end{align}
Thus, the probabilistic constraint~\eqref{eq:QoS3-probabilisic} can be assured by asking for the following constraint involving \PSD\ of $Z_\ell$ to be satisfied:
\begin{align} \label{eq:exampSuffCond-2}
\frac{1}{2\pi}\int_{-\pi}^{\pi} S_{Z_\ell}(\discFreq) d\discFreq \leq b_\ell =: c_\ell^2\varepsilon_\ell.
\end{align}
If the dynamic system $\G_{\ell}$ relating the input $\tilde{P}$ and output $Z_\ell$ were a linear time invariant system, then \eqref{eq:exampSuffCond-2} can be translated to a constraint on the power deviation:
\begin{align} \label{eq:exampSuffCond-2}
\frac{1}{2\pi}\int_{-\pi}^{\pi} S_{\tilde{P}}(\discFreq) |G_{\ell}(e^{j\discFreq})|^2d\discFreq \leq b_\ell
\end{align}
where $G_\ell(e^{j\discFreq})$ is the frequency response of the dynamic system $\G_\ell$. In the general case with $m$ constraints, the constraint set for the \PSD\ $S_{\tilde{P}}$ is
\begin{align} \label{eq:consSetPastWork}
\setOfPSD \triangleq \left\{ S_{\tilde{P}} \ \bigg\vert  \ \frac{1}{2\pi}\int_{-\pi}^{\pi}S_{Z_\ell}(\discFreq;S_{\tilde{P}})d\discFreq \leq b_\ell, \quad \ell=1,\dots,m \right\}.
\end{align}
The notation $S_{Z_\ell}(\discFreq;S_{\tilde{P}})$ above is used to emphasize that the \PSD\ of the signal $Z_\ell$ is a function of  $S_{\tilde{P}}$, the \PSD\ of $\tilde{P}[k]$. The function can be arbitrarily complex when the dynamic models $\G_\ell$ are nonlinear.

As long as the \PSD\ of the demand deviation $\tilde{P}[k]$ belong to the set \setOfPSD, each of the $\ell$ probabilistic QoS constraints - such as QoS 1-4 described in Section~\ref{sec:indLoadModel} - holds. Thus, the set \setOfPSD\ also represents the demand deviation capacity of the flexible load. 
\def\H{\ensuremath{\mathcal{H}}}

Now, denote by $\H$ the set of \PSD s defined over $[-\pi,\pi)$, and define the function $B_\ell: \H \to \R^+$ as
\begin{align}
\label{eq:Bbardef-general}
\bar{B}_\ell(S_{\tilde{P}}) = \frac{1}{2\pi}\int_{-\pi}^\pi S_{Z_\ell}(\Omega; S_{\tilde{P}})d\Omega
\end{align}
and the array $\bar{B}(S_{\tilde{P}})=[\bar{B}_1(S_{\tilde{P}}),\dots, \bar{B}_m(S_{\tilde{P}})]^T \in \R^m$. By denoting
$\vecOfQoS = [	\varepsilon_1c^2_1,\dots,	\varepsilon_mc^2_m]^T \in \R^m$ the constraint $S_{\tilde P} \in \setOfPSD$ can be represented as
\begin{align}\label{eq:B-ineq}
\bar{B}(S_{\tilde P}) \leq \vecOfQoS.
\end{align}

Now consider the following optimization problem:
\begin{align}\label{prob:optProbDiscTime}
\begin{split}
&\min_{S} \quad \frac{1}{2\pi}\int_{-\pi}^{\pi}\left(S(\discFreq) -S^{\loadDem}(\discFreq)\right)^2d\discFreq  \\ 
&\text{s.t.} \quad \bar{B}(S_{\tilde{P}}) \leq \vecOfQoS \ \text{ and} \ S(\Omega) \geq 0 \; \quad \forall \ \Omega \in [-\pi, \pi),    
\end{split}
\end{align}
where $S^{\loadDem}(\discFreq)$ is the spectral density of the stochastic process that generates the reference signals from the BA. How to determine $S^{\loadDem}(\discFreq)$ is discussed in the next section. \emph{We define the capacity of the flexible load as the solution $S^*_{\tilde P}$ of the optimization problem~\eqref{prob:optProbDiscTime}.}

\subsubsection{BA's spectral needs} \label{sec:BAspecNeed}
The \emph{total} needs of the BA is encapsulated by the \PSD\ of the net demand signal, an example of which is shown in Figure~\ref{fig:Resource_Power}. With historical data, a BA can estimate the \PSD\ of the net demand signal, which we denote as $S^{\netDem}(\discFreq)$. Any well posed estimation technique can be applied. All controllable resources, including generators, flywheels, batteries, and flexible loads, together have to supply $S^{\netDem}(\discFreq)$. To determine solely the portion of $S^{\netDem}(\discFreq)$ that flexible loads should contribute to we ``filter'' $S^{\netDem}(\discFreq)$. That is, with $F(e^{j\discFreq})$ an appropriate filter we have 
\begin{align} \label{eq:filtNetDem}
S^{\loadDem}(\discFreq) = \left|F(e^{j\discFreq})\right|^2S^{\netDem}(\discFreq).
\end{align}
The quantity $S^{\loadDem}$ is the frequency domain analog of the reference signal $r[k]$ that will be asked from the loads, and will be referred to as the \emph{reference \PSD} in the sequel.


\subsection{Linear time invariant (model-based) case: prior work} \label{sec:limPastWork}
Now we consider the scenario when the $\G_\ell$'s are LTI systems. Recall that $S_{Z_\ell}(\discFreq; \tilde{P})$ in the definition of $B(\cdot)$ in~\eqref{eq:Bbardef-general} is the \PSD\ of $Z_\ell[k]$, which in turn is the output of the system $\G_\ell$ when driven by an input signal whose \PSD\ is $S_{\tilde P}(\discFreq)$. Since the system $\G_\ell$ is LTI with frequency response $G_\ell(e^{j\discFreq})$, it follows from Prop.~\ref{prop:psdbackground} that
\begin{align}
  S_{Z_\ell}(\discFreq) = |G_\ell(e^{j\discFreq})|^2 S_{\tilde P}(\discFreq),
\end{align}
and plugging it in~\eqref{eq:Bbardef-general} we get
\begin{align}
  \bar{B}_\ell(S_{\tilde P}) = \frac{1}{2\pi}\int_{-\pi}^{\pi} |G_\ell(e^{j\discFreq})|^2 S_{\tilde P}(\discFreq)d\discFreq.
\end{align}
The constraints in Problem~\eqref{prob:galRelProbDiscTime} are thus linear in the decision variable $S_{\tilde{P}}(\discFreq)$. Since the objective is quadratic in the decision variable, the problem is a quadratic problem, although infinite dimensional.

\begin{remark}\label{rem:finiteDimOpt}
 The problem~\eqref{prob:galRelProbDiscTime} can be reduced to a tractable finite dimensional optimization problem by discretizing the continuous frequency $\discFreq$ into $N$ points on the unit circle. The decision vector of the optimization problem becomes $N$. The resulting problem is a finite dimensional quadratic program (QP) that can be efficiently solved using readily available NLP solders. \emph{In all such problems in the rest of the paper that involve functions of continuous frequency $\discFreq$ over $[-\pi, \pi]$, we assume that such a discretization is done to convert the problem to a finite dimensional problem.} \hfill $\square$
\end{remark}

The finite dimensional QP alluded to in Remark~\ref{rem:finiteDimOpt} is the problem posed and solved in our prior work~\cite{CoffmanSpectral_ACC:2020}. Thus, the optimization problem needed to characterize capacity is fairly straightforward to solve, \emph{as long as the models $\G_\ell$'s are LTI and the model parameters are known}. There are two weaknesses. The first is that a linear model may not be appropriate for certain types of flexible loads. The second is that even if a LTI model is sufficiently accurate, obtaining the model parameters is not an easy task. Take the LTI model~\eqref{eq:storage-LTI} of temperature deviation in a building. This equation alone is actually quite merited for this particular application, and there is a plethora of work spanning back to the 1980's~\cite{Rabl:1988} on using ODEs of this form to model the dynamics in certain flexible loads. These works almost solely focus on \emph{estimating} the parameters of the model such as~\eqref{eq:storage-LTI}. These parameters are challenging to estimate. 
Despite this, many current capacity characterizations explicitly depend on the parameters such as $R$ and $C$.

\section{Proposed Data Driven Method} \label{sec:sepCase}
The goal of this section will be to develop an algorithm that can solve the problem~\eqref{prob:optProbDiscTime} using data that can come from experiments or simulations, but without requiring (i) that the underlying systems $\G_\ell$'s are LTI and (ii) any knowledge of the models $\G_\ell$'s. Only a  simulator that can simulate $\G_\ell$'s for various inputs is needed. 

To facilitate our algorithm, we first elect a function approximation architecture for the decision variable $S$ in the optimization problem~\eqref{prob:optProbDiscTime}. With our form of function approximation, we show how to obtain an estimate of all of the ingredients needed to solve~\eqref{prob:optProbDiscTime} with solely data. 

\subsection{Function approximation}
We consider linear function approximations, that is, we approximate the decision variable $S$ in~\eqref{prob:optProbDiscTime} through
\begin{align}
	S^{\theta}(\discFreq) = \sum_{i=1}^{d}\psi_i(\discFreq)\theta_i = \Psi^T(\discFreq)\theta,
\end{align}
where each \emph{basis} $\psi_i(\discFreq)$ is a \PSD, and $\theta \geq 0$. The number of basis functions, $d$, is a design choice. \emph{We use $\Psi^T\theta$ to denote the entire trajectory $\{\Psi^T(\discFreq )\theta)\}_{\discFreq=-\pi}^{\discFreq=\pi}$}.

We then transform the optimization problem~\eqref{prob:optProbDiscTime} over $S$ to one over the finite dimensional vector $\theta \in R^d$. The problem~\eqref{prob:optProbDiscTime} is transformed to a finite dimensional non-linear program (NLP): 
\begin{align} \label{prob:galRelProbDiscTime}
  \begin{split}
\theta^* = \arg\min_{\theta} \quad &\frac{1}{2\pi}\int_{-\pi}^{\pi}\left(\Psi^T(\discFreq)\theta -S^{\loadDem}(\discFreq)\right)^2d\discFreq  \\ 
& \text{s.t.} \quad B(\theta) \leq \vecOfQoS ,\quad \text{and} \quad \theta \geq 0,    
  \end{split}
\end{align}
where $B(\theta) := \bar{B}(\Psi^T\theta)$, where $\bar{B}(\cdot)$ is defined in~\eqref{eq:Bbardef-general}. Since $\psi_i(\discFreq)\geq 0$ for each $i$, requiring $\theta \geq 0$ ensures that $\Psi^T(\discFreq)\theta$ satisfies the properties of \PSD s (non-negativity and even) and so the search is limited to \PSD s, and the solution obtained by solving the problem~\eqref{prob:galRelProbDiscTime}, $\Psi^T(\discFreq)\theta^*$, is guaranteed to be a \PSD.

\subsection{Estimating $B(\theta)$ from data} \label{sec:estBthetaData}
The method we propose for estimating $B_\ell(\theta)$ for a given $\theta$ and a fixed $\ell$ is given below. It is then repeated for $\ell=1,\dots,m$ to obtain $B(\theta)$
\begin{enumerate}
\item Generate samples of the $\ell$-th QoS signal, $Z_\ell[k]$, when power deviation $\tilde{P}[k]$ has \PSD\ $\Psi^T(\discFreq)\theta$. This is done in two steps:
\begin{enumerate}
\item \underline{Input generation:} For each $i$ ($i=1,\dots,d$) generate a colored noise sequence $\varphi_i[k]$ with \PSD\ $\theta_i\Psi_i(\Omega)$. (This can be done in many ways. One possibility is to perform a spectral factorization of $\Psi_i$ to obtain a filter $H(e^{j\Omega})$ so that $|H(e^{j\Omega})|^2 = \Psi_i(\Omega)$. Passing a zero mean unit variance white noise through will generate a WSS process with \PSD\ $\Psi_i(\Omega)$ due to Prop.~\ref{prop:psdbackground}. Multiplying this sequence with $\sqrt{\theta_i}$ will produce the desired sequence $\varphi[k]$. Another method is to take the inverse (discrete time) Fourier transform of the \PSD\ multiplied pointwise (in the frequency domain) by random phase.)
\item \underline{Output generation:} Use a simulator of the system $\G_\ell$ to generate $Z_{i,\ell}[k]$ by using the input $\varphi_i[k]$, for each $i$, and then sum over $i$ to obtain $Z_{\ell}[k] := \sum_{i=1}^d Z_{i,\ell}[k]$. (Because the same simulator is used for each $i$, we have $Z_{\ell}[k] = \G_\ell(u)[k]$ where $u[k] = \sum_{i} \varphi_i[k]$. Because the processes $\varphi_i[k]$ and $\varphi_j[k]$ are uncorrelated for $i\neq j$, the \PSD\ of $u $ is the sum of the \PSD s of $\varphi_{i}[k]$'s, which is equal to $\Psi^T(\discFreq)\theta$ by design. Thus, the \PSD\ of $Z_{\ell}[k]$ is \PSD\ of the output of $\G_\ell$ with an input whose \PSD\ is $\Psi^T(\discFreq)\theta$. In other words, the \PSD\ of $Z_{\ell}[k]$ is $S_{Z_\ell}(\discFreq; \Psi^T(\discFreq)\theta)$.)
 \end{enumerate}
\item Estimate the function value $B_\ell(\theta)$ from the samples $Z_\ell[k]$ by utilizing the Wiener-Khinchin  theorem. Namely,
  \begin{align}\label{eq:Bhat-basic}
    &\hat{S}_{Z_\ell}(\discFreq; \Psi^T\theta) = \hat{\Expect}\Big[\frac{1}{N}\Big| \sum_{k=1}^N Z_{\ell}[k]e^{-j \Omega k}\Big|^2\Big], \\
    &\hat{B}_{\ell}(\theta) = \frac{1}{2\pi}\int_{-\pi}^{\pi}\hat{S}_{Z_\ell}(\discFreq; \Psi^T\theta) d\discFreq
  \end{align}
  where $\hat{\Expect}[\cdot]$ is a shorthand for an estimate of the expectation $\Expect[\cdot]$. In particular, the estimate is obtained by performing multiple simulations, computing the quantity inside the square braces in the right hand side of~\eqref{eq:Bhat-basic} in each simulatin, and averaging over those simulations. 
\end{enumerate}

\begin{figure}[ht]
  \centering
  \includegraphics[width=0.9\columnwidth]{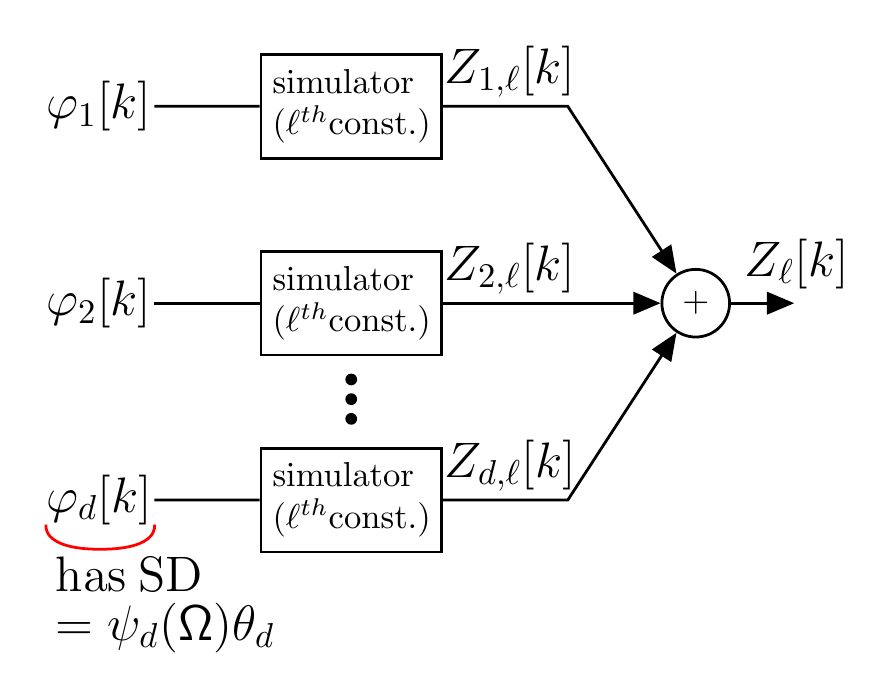}
  \caption{Generating the $\ell$-th QoS signal $Z_\ell[k]$.}
  \label{fig:algo-mainsteps}
  \vspace{-0.3 cm}
\end{figure}

A graphical illustration of the data generation step of the algorithm is shown in Figure~\ref{fig:algo-mainsteps}.
To completely specify the problem~\eqref{prob:galRelProbDiscTime}, the quantity $S^{\loadDem}(\discFreq)$ is required. As mentioned in section~\ref{sec:BAspecNeed}, it can be estimated using net load data. 

The problem~\eqref{prob:galRelProbDiscTime} is a finite dimensional NLP, with a $d$-dimensional decision vector $\theta$. However, standard NLP solvers cannot be easily used in the general nonlinear case.  Since the \PSD\ of the output $Z_\ell[k]$ can be an arbitrarily complicated function of $\theta$, the gradient and/or Hessian of the function $B(\theta)$ is not readily available. They can be numerically estimated, or a gradient free method can be used. In the special case when the systems $\G_\ell$'s are LTI, the problem becomes a QP, a convex problem that can be easily solved. We describe this special case in Section~\ref{sec:opt-LTI}.

\begin{remark}
	Key in our ability to remove dependence on the model knowledge is the requirement that each basis function $\Psi_i$ is in fact a \PSD. Without this form of function approximation it may be difficult to develop a truly model free form of the problem~\eqref{prob:optProbDiscTime}. As we will discuss next, this model free dependence rids us of the limitations of the past work discussed in Section~\ref{sec:limPastWork}.
\end{remark}

\subsection{The LTI case}\label{sec:opt-LTI}
Since $S_{\tilde{P}} = \Psi^T \theta$, it follows from~\eqref{eq:Bbardef-general} that
\begin{align}
  \label{eq:3}
  \bar{B}_\ell(\theta) = B_\ell(\Psi^T \theta) = \frac{1}{2\pi}\int_{-\pi}^{\pi}S_{Z_\ell}(\discFreq; \Psi^T \theta) d\discFreq.
\end{align}
Recall that $S_{Z_\ell}(\discFreq; \Psi^T \theta)$ is the \PSD\ of $Z_\ell[k]$ which is the output of system $\G_\ell$ when driven by a signal whose \PSD\ is $\Psi^T(\discFreq) \theta$. Since the system $\G_\ell$ is LTI with frequency response $G_\ell(e^{j\discFreq})$, it follows from Prop~\ref{prop:psdbackground} that
\begin{align}
  \label{eq:4}
  S_{Z_\ell}(\discFreq) = |G_\ell(e^{j\discFreq})|^2 \Psi^T(\discFreq) \theta.
\end{align}
Plugging it back in~\eqref{eq:3} we get
\begin{align}
  \label{eq:5}
  \bar{B}_\ell(\theta) = \left[\frac{1}{2\pi}\int_{-\pi}^{\pi} |G_\ell(e^{j\discFreq})|^2 \Psi^T(\discFreq) d\discFreq \right] \theta
\end{align}
Stacking these $\bar{B}_\ell(\theta)$'s, we obtain
\begin{align}
  \label{eq:5}
    \bar{B}(\theta)  = B\theta
\end{align}
where $B \in \mathbb{R}^{m\times d}$ is
\begin{align} \label{eq:Bmat}
B & = \frac{1}{2\pi}\int_{-\pi}^{\pi}\vecOfG(\discFreq)\Psi^T(\discFreq)d\discFreq,\\
  \vecOfG(\discFreq) & = [|G_1(e^{j\discFreq})|^2,\dots, |G_m(e^{j\discFreq})|^2]^T \notag
\end{align}
The objective function~\eqref{prob:galRelProbDiscTime} can be expressed as
\begin{align}
	f(\theta) = \theta^TA\theta + C\theta + d,
\end{align}
where
\begin{align}
	A &= \frac{1}{2\pi}\int_{-\pi}^{\pi}A(\discFreq)d\discFreq, \ \text{with} \ A(\discFreq) = \Psi(\discFreq)\Psi^T(\discFreq), \\
	C &= \frac{1}{2\pi}\int_{-\pi}^{\pi}C(\discFreq)d\discFreq, \ \text{with} \ C(\discFreq) = \Psi(\discFreq)S^{\loadDem}(\discFreq), \\
	d &= \frac{1}{2\pi}\int_{-\pi}^{\pi}d(\discFreq)d\discFreq, \ \text{with} \ d(\discFreq) = \big(S^{\loadDem}(\discFreq)\big)^2.
\end{align}
To estimate the quantities $C$ and $d$ above, the quantity $S^{\loadDem}(\discFreq)$ is required. As mentioned in section~\ref{sec:BAspecNeed}, it can be estimated using net load data. 
\begin{align} \label{eq:fHatEst}
	f(\theta) = \theta^TA\theta + C\theta + d.
\end{align}
The optimization problem~\eqref{prob:galRelProbDiscTime} now becomes the following $d$-dimensional quadratic program (QP)
\begin{align} \label{eq:appQuadProg}
{\theta}^* = \arg\min_{\theta\geq 0} \ f(\theta), \quad
 \text{s.t.} \quad B\theta \leq \vecOfQoS.	
\end{align}
To reiterate, in order to compute $f(\theta)$ and $B$, no model knowledge is required, only access to a simulator.

\ifx 0
\section{Data Driven algorithm: Non separable case} \label{sec:nonSepCase}
The problem becomes more challenging to solve when the constraints are not linear. The reason for this is that linearity allowed for interpretations such as~\eqref{eq:exampNonParaEst}. This interpretation then sparked the data driven algorithm given in Section~\ref{sec:dataDrivenAlgor}. However, for example, if the LHS of one of constraints is of the form
\begin{align}
	\frac{1}{2\pi}\int_{-\pi}^{\pi}H_\ell\big(\discFreq,\Psi^T(\discFreq)\theta\big)d\discFreq, 
\end{align}
where $H_\ell(\cdot,\cdot)$ is a \PSD\ of the $\ell^{th}$ output, then the situation becomes challenging. It's dependence on the power deviation \PSD\ is described in the most general sense through its second argument. In this setting, there will most likely not be a factorization of the form 
\begin{align}
	H_\ell(\discFreq,\Psi^T(\discFreq)\theta) = G_\ell(\discFreq)\Psi^T(\discFreq)\theta.
\end{align}
Hence, the algorithm described in Section~\ref{sec:dataDrivenAlgor} is not applicable here. Before proceeding, we highlight that assuming the constraints are linear or not requires some knowledge of the model structure. For example, the storage constraint described in section~\ref{sec:unkModParam} is of linear form whereas the MISO system described in section~\ref{sec:misoSysConst} will yield a constraint of non-linear form. However, in principle, the data driven nature of the algorithm allows one to model the constraints either as linear or non-linear. Although, modeling the constraint described in~\ref{sec:misoSysConst} as linear will likely yield poor results. 

\subsection{A SQP Algorithm}
The main challenge in non separable constraints is that one cannot simply first estimate the matrix $B$ appearing in~\eqref{prob:galRelProbDiscTime} and then use this to solve for $\theta^*$. This is because, in the non-separable case, the matrix $B$ will also depend on $\theta$. To overcome this, we will present a sequential quadratic programming (SQP) algorithm. We elect an algorithm of this form for two reasons: (i) it will still preserve the data driven nature of the problem and (ii) it is a common approach to solving optimization problems with non-linear constraints. 

To clarify, the problem we are solving is the following
\begin{align} \label{prob:galRelProbNonLin}
\theta^* = \arg \min_{\theta \geq 0} \quad &\frac{1}{2\pi}\int_{-\pi}^{\pi}\left(\Psi^T(\discFreq)\theta -S^{\loadDem}(\discFreq)\right)^2d\discFreq \\ 
& \text{s.t.} \quad B^{L}\theta \leq \vecOfQoS^L  \ \leftrightarrow \lambda^L \\ \label{eq:nonLinCons}
&\qquad B^{NL}(\theta) \leq \vecOfQoS^{NL} \ \leftrightarrow \lambda^{NL}.
\end{align}
The non-linear constraints are then captured in~\eqref{eq:nonLinCons}. We denote the Lagrangian of the problem~\eqref{prob:galRelProbNonLin}:
\begin{align}
	L(\theta,\lambda) = f(\theta) + \lambda^Tg(\theta),
\end{align}
where $\lambda = [(\lambda^L)^T, (\lambda^{NL})^T]^T$, $f$ is the objective function in~\eqref{prob:galRelProbNonLin} and $g$ is the vector of all of the constraint functions
\begin{align}
	g(\theta) = \begin{bmatrix}
	B^L\theta - \vecOfQoS^L \\
	B^{NL}(\theta) - \vecOfQoS^{NL}
	\end{bmatrix}.
\end{align} 
 The SQP approach to solve~\eqref{prob:galRelProbNonLin} is then an iteration of the form
\begin{align}
	\theta_{k+1} = \theta_k + \alpha_k d_k,
\end{align} 
where $\alpha_k$ is a step size and $d_k$ is a search direction. The name SQP comes from the sequential procedure to obtain the search direction $d_k$:
\begin{align}
	d_k = \arg\min_d \ &\nabla f(\theta_k)^Td + \frac{1}{2}d^T \Big(\nabla^2 L(\theta_k,\lambda_k)\Big)d \\
	&\text{s.t.} \quad g(\theta_k) + \nabla g(\theta_k)d \leq 0,
\end{align}
where $\nabla (\cdot)$ denotes gradient with respect to $\theta$.
\fi
\section{Numerical Experiments} \label{sec:numExp}
A numerical example of using the proposed data driven method to determine the capacity is illustrated in this section. The flexible loads considered are a collection of commercial building HVAC systems. We consider a homogeneous collection in this preliminary work. 
Each HVAC system in the collection has QoS 1-4 listed in Section~\ref{sec:indLoadModel}. The parameters are displayed in Table~\ref{tab:simParams_bldg}, and are chosen so that the HVAC systems are representative of those in large commercial buildings (hence the large superscripts in Table~\ref{tab:simParams_bldg}).

We first validate the proposed method by implementing it in the scenario in which the solution is known: when the models involved are LTI and are known. The solution is computed using the method from our prior work~\cite{CoffmanSpectral_ACC:2020}. Next, we apply the method to data from a simulation that uses a non-linear model.

All relevant simulation parameters, if not specified otherwise, can be found in Table~\ref{tab:simParams_bldg}. Note that the method does \emph{not} have access to the $R$, $C$, and $\eta_{0}$ parameter values.

\begin{table}
	\centering
	\caption{Simulation parameters}
	\label{tab:simParams_bldg}
	\begin{tabular}{|| l| c |c|| l |c |c||}
		Par.    & Unit      & Value  & Par.          & Unit         & Value \\
		$R$   & $^{\circ}$C$/$kW        & 8      & $C$  & kWh$/^{\circ}$C     & 22\\
		$T$   & hours        & 5    & $\{\epsilon_i\}_{i=1}^4$   &N/A& 0.05 \\
		$c_{1}$ &kW& 40   & $c_{2}$  & kW     & 8\\
		$c_{3}$   & $^\circ$C       & 1    & $c_4$   &kWh& 8 \\
		$\eta_{0}$   & N/A       & 3.5    & $T_a$   &$^\circ$C& 30 
		\\
		$\delta$   & sec.       & 20    & $\dot{q}_{\text{int}}$   &kW& 0 
	\end{tabular}
\end{table}

\subsection{BA's spectral needs}\label{sec:obtain_Ref_PSD} 
\begin{figure}
	\centering
	\includegraphics[width=\columnwidth]{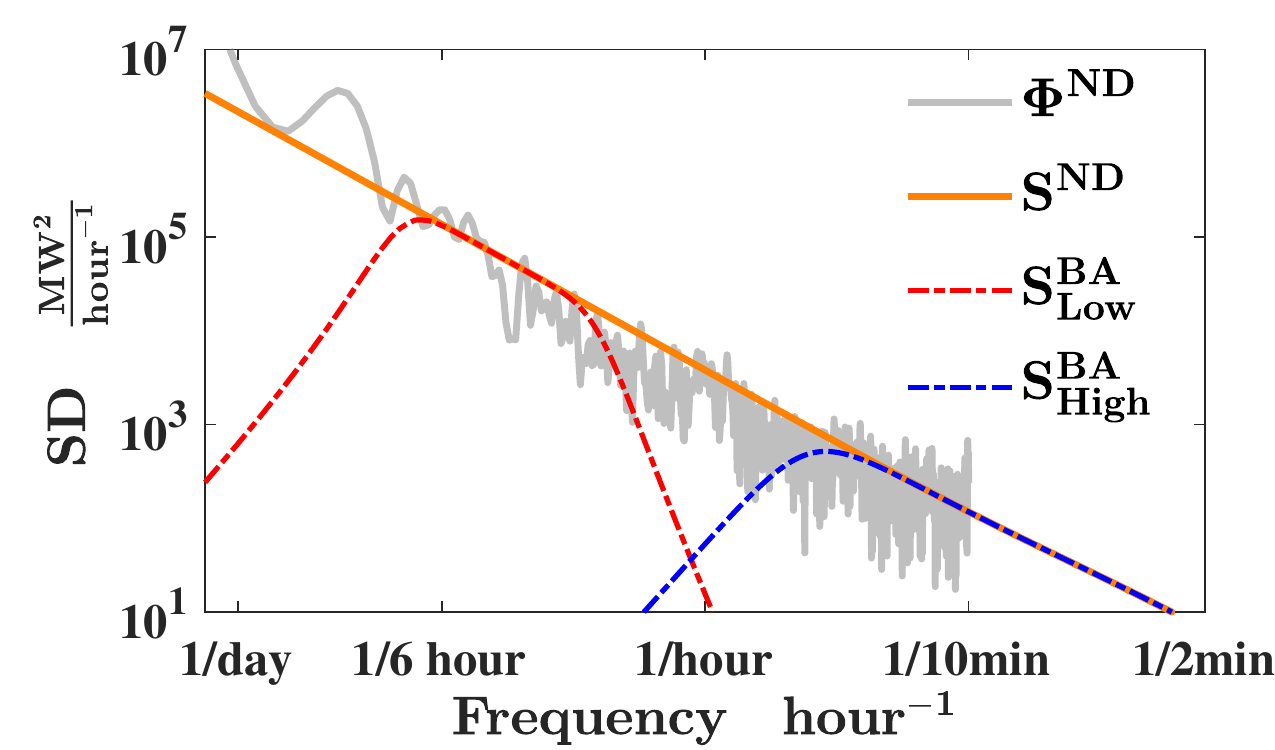}
	\caption{Empirical net demand \PSD, modeled \PSD\ for BPA's net demand, and the two reference \PSD's for the high and low frequency passband.}
	\label{fig:NT_PSD}
\end{figure}

The net demand data is collected from BPA (a BA in the pacific northwest United States). The empirical \PSD\ of the net demand is determined using the method described in Section~\ref{sec:BAspecNeed}. We then fit an ARMA(2,1) model to the empirically estimated \PSD. Since the estimate $\Phi^{\netDem}$ will cap out at the Nyquist frequency $1/10$min, we extrapolate the net demand \PSD\ to the higher frequencies. The empirical \PSD\ (denoted $\Phi^{\netDem}$) and the extrapolated net demand \PSD\ (denoted $S^{\netDem}$) are shown in Figure~\ref{fig:NT_PSD}. 

We then choose two passbands to filter $S^{\netDem}$: (i) a low passband [$1/6$,$1/2$] (1/hour) and (ii)  a high passband [$1/30$,$1$] (1/min). The results of ``filtering'' (see eq.~\eqref{eq:filtNetDem}) $S^{\netDem}$ are also shown in Figure~\ref{fig:NT_PSD}. The low passband \PSD\ is termed $S^{\loadDem}_{\textLow}$ and roughly corresponds to the region for TCLs in Figure~\ref{fig:Resource_Power}. The high passband \PSD\ is termed $S^{\loadDem}_{\textHigh}$ and roughly corresponds to the region for HVAC systems in Figure~\ref{fig:Resource_Power}.

\subsection{Method Evaluation - LTI case} \label{subsec:resHomoLoad}
In this section we compare the data-driven method for the LTI case (Section~\ref{sec:opt-LTI}) with to solving the quadratic program~\eqref{eq:appQuadProg} with full model knowledge, both to obtain $\theta^*$. In both cases we use CVX~\cite{cvx} to solve the QP. We use the following basis \PSD's
\begin{align}
	\psi_i(\discFreq) = \begin{cases}
	1, & \text{if} \ \discFreq \in [\hat{\discFreq}_{i-1},\hat{\discFreq}_{i}). \\
	0, & \text{otherwise}.
	\end{cases}
\end{align}
for $1\leq i \leq d$. The set of points $\{\hat{\discFreq}_{i}\}_{i=1}^d$ is a subset of the linearly spaced discrete frequency points on the unit circle. We consider  $\numLoads = 2000$ large commercial buildings as one large flexible load. The idea is to illustrate how much of the grids needs can be met by the collection. To do this, the two reference \PSD s obtained from the previous section are projected onto the \emph{same} ensemble constraint set. 

\begin{figure}
	\centering
	\includegraphics[width=1\columnwidth]{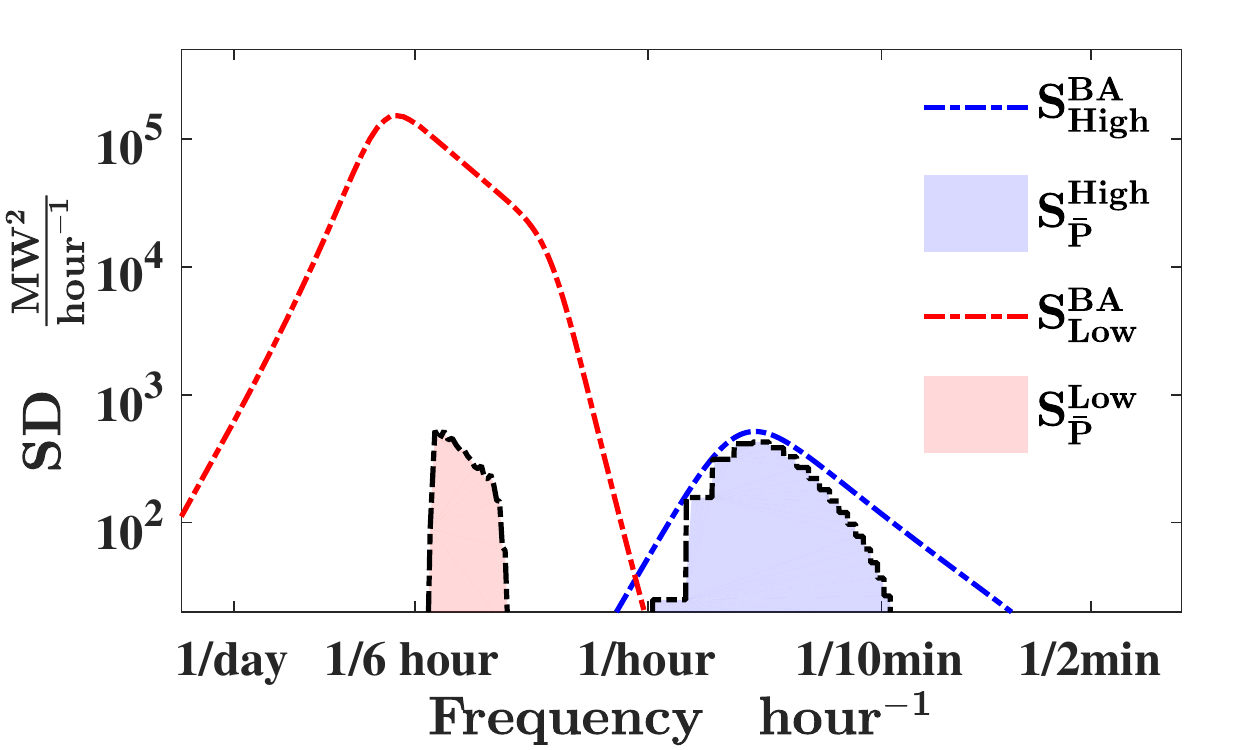}
	\caption{LTI case: The two reference \PSD s and the corresponding capacity \PSD s (boundary of the shaded regions) obtained from the proposed method for a homogeneous collection of $\numLoads=2000$ loads. Black dashed lines represent model based solution.}
	\label{fig:Ref_PSD_case1}
\end{figure}

The results of this are shown in Figure~\ref{fig:Ref_PSD_case1}, where the black dashed lines represent the model based solution. The two \PSD's are nearly identical. 

\ifx 0
\begin{table}
	\centering
	\def\arraystretch{1.4}%
	\caption{Equivalent Battery Capacity}
	\label{tab:equivBattCap}
	\begin{tabular}{||l || c |c||}
		Num. of Loads & High Passband & Low Passband \\
		\hline
		3000 & 120 MW \& 12.6 MWh & 19 MW \& 15 MWh \\
		6000 & 120 MW \& 12.6 MWh & 37 MW \& 30 MWh \\
		9000 & 120 MW \& 12.6 MWh & 55 MW \& 45 MWh \\
		12000 & 120 MW \& 12.6 MWh & 73 MW \& 60 MWh \\
		15000 & 120 MW \& 12.6 MWh & 90 MW \& 75 MWh
	\end{tabular}
\end{table}
\fi

\subsection{Method Evaluation: Non-linear case} \label{sec:nonLinConNumExp}
In this section we use the (discrete-time version of the) nonlinear dynamic model~\eqref{eq:ODE-nonlinear} relating power deviation to indoor temperature. In this scenario, all of the parameter values (except for now we have $\eta[k]$) remain the same as in the previous scenario and are given in Table~\ref{tab:simParams_bldg}. The same basis \PSD's as in the previous scenario are also used here. We elect the value $\alpha$ appearing in~\eqref{eq:nonLinCOPMod} as $\alpha_1 = 0.15$ and $\alpha_2 = 1.175$. 

\begin{figure}
	\centering
	\includegraphics[width=1\columnwidth]{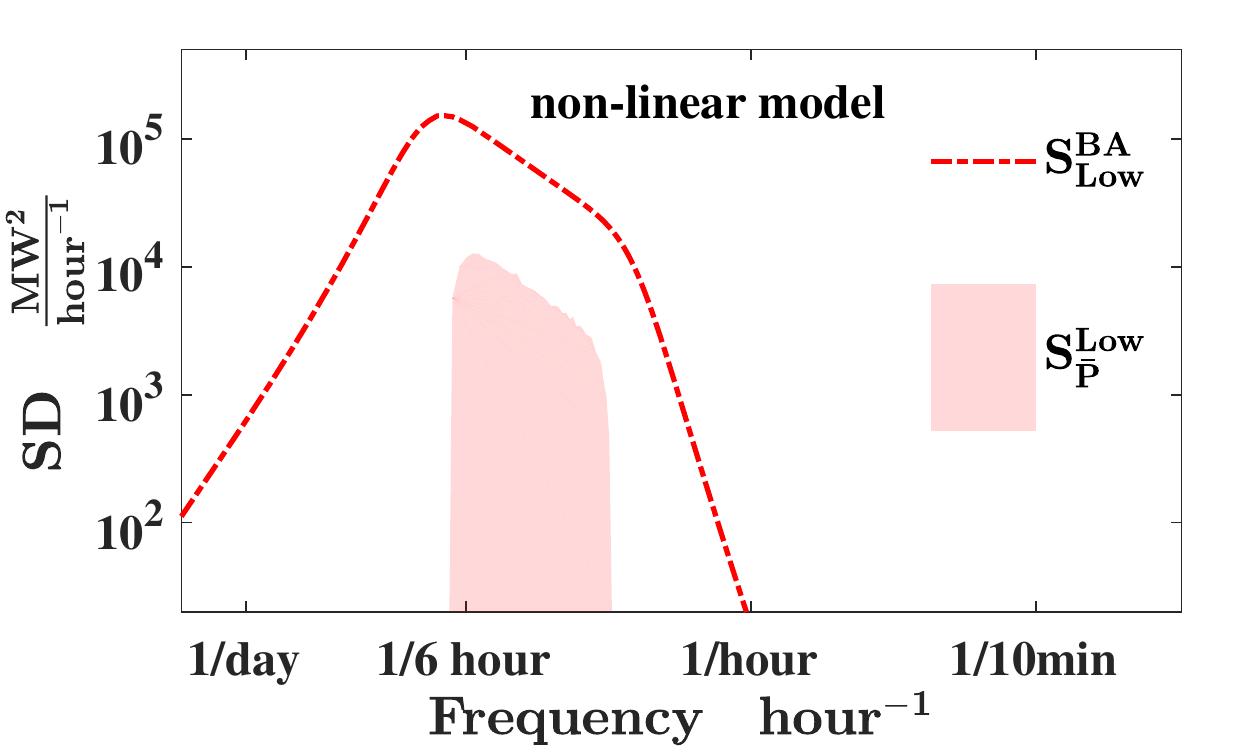}
	\caption{Non linear case: The capacity (and reference) \PSD\ for $\numLoads = 15000$  flexible loads, each with a non-linear model. }
	\label{fig:nonLinPSDresult}
\end{figure}

The proposed method is applied to simulation data from the non-linear system. The reference \PSD\ is $S^{\loadDem}_{\textLow}$ and $\numLoads = 15000$. It is possible to use a standard NLP solver to solve~\eqref{prob:galRelProbDiscTime}, however we obtained positive results by applying the data driven algorithm in Section~\ref{sec:opt-LTI} to data collected from the nonlinear system. We feel this result is more interesting, and choose to show it here instead of results from the NLP solver. We emphasize that the method does not know anything about the model, it only uses simulation data. The results - the solution \PSD\ - is shown in Figure~\ref{fig:nonLinPSDresult}.

To verify that the solution provided by the method is meaningful, we generate power deviation trajectories with \PSD\ equal to the solution \PSD. Then the corresponding demand trajectories are used to simulate the model again using the simulator to compute the resulting temperature deviations. One such temperature trajectory is shown in  Figure~\ref{fig:nonLinPSDresult}. 
We see from the figure that temperature is maintained within bounds during the time interval shown, indicating that the power deviation signal is within the capacity of the load.

\begin{figure}
	\centering
	\includegraphics[width=1\columnwidth]{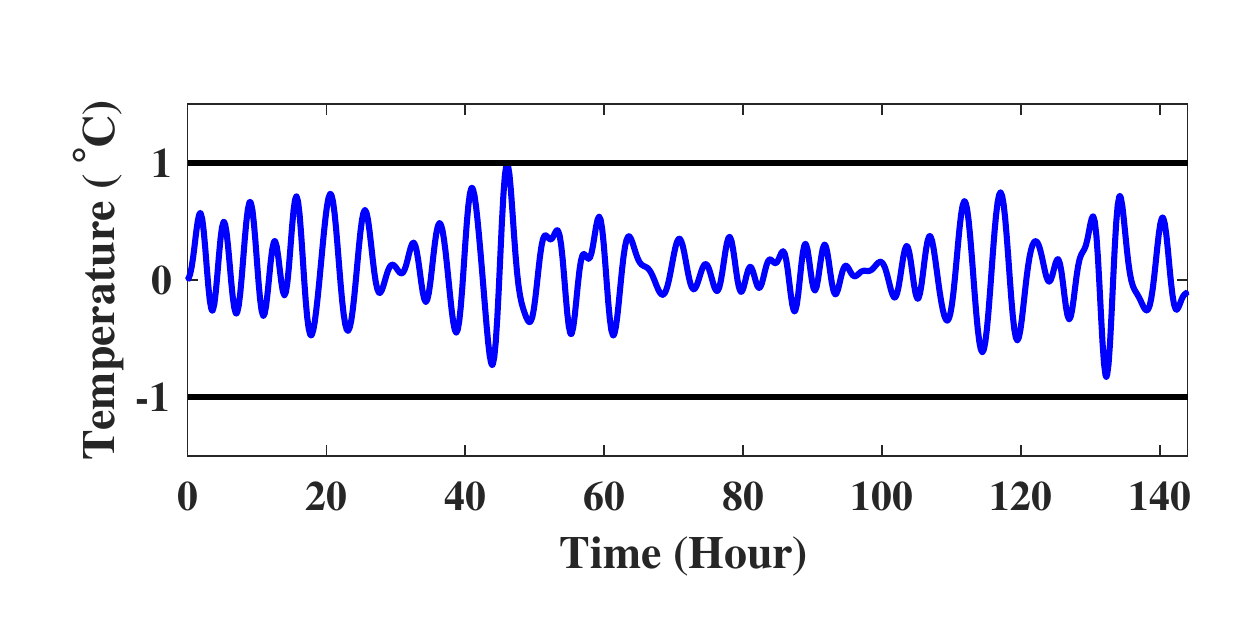}
	\caption{Sample path of a flexible HVAC load's temperature deviation from setpoint (evolves according to~\eqref{eq:ODE-nonlinear}). The black lines represent the QoS constraint.
        }
	\label{fig:nonLinTimeSeries}
\end{figure}

\section{Conclusion} \label{sec:conc}
We presented a data driven method to estimate the capacity of flexible load(s) as the optimal spectral density of demand deviation. Optimal here refers to being close to what the power grid needs. The methd builds on our prior work~\cite{CoffmanSpectral_ACC:2020} which was model-based and was limited to LTI models. The method proposed here is also applicable to nonlinear dynamics, and more importantly, it does not need model knowledge. It only needs access to a simulator (or measurements of relevant data). The core of the algorithm is a function approximation architecture with basis functions that are chosen to be spectral densities. In simulations, our proposed data-driven method is validated against the model knowledge scenario; the results are positive.

Solving the projection problem in the nonlinear dynamics case is not trivial since symbolic derivatives with respect to the decision variables is not possible. This aspect of the method has room for improvement. Other avenues for future work include leveraging the data driven framework to: (i) estimate the capacity of heterogeneous ensembles of flexible loads and (ii) estimate the capacity under time varying weather conditions/disturbances.

\end{document}